\documentclass[preprint,aps,showpacs,showkeys,preprintnumbers,amsmath,superscriptaddress,floatfix,amssymb]{revtex4}

\usepackage{graphicx}
\usepackage{dcolumn}
\usepackage{bm}

\begin{document}

\preprint{}

\title{d-$\alpha$ correlation functions and collective motion in 
Xe+Au collisions at E/A=50 MeV}


\author{G.~Verde}
\affiliation{ Istituto Nazionale di Fisica Nucleare, Sezione di Catania, 64 Via Santa Sofia, I-95123 Catania, Italy}
\author{P.~Danielewicz}
\affiliation{ National Superconducting Cyclotron Laboratory and Department of Physics and Astronomy, Michigan State University, East Lansing, MI 48824, USA}
\author{W.G.~Lynch}
\affiliation{ National Superconducting Cyclotron Laboratory and Department of Physics and Astronomy, Michigan State University, East Lansing, MI 48824, USA}
\author{C.F.~Chan}
\affiliation{ National Superconducting Cyclotron Laboratory and Department of Physics and Astronomy, Michigan State University, East Lansing, MI 48824, USA}
\author{C.K.~Gelbke}
\affiliation{ National Superconducting Cyclotron Laboratory and Department of Physics and Astronomy, Michigan State University, East Lansing, MI 48824, USA}
\author{T.X.~Liu}
\affiliation{ National Superconducting Cyclotron Laboratory and Department of Physics and Astronomy, Michigan State University, East Lansing, MI 48824, USA}
\author{X.D.~Liu}
\affiliation{ National Superconducting Cyclotron Laboratory and Department of Physics and Astronomy, Michigan State University, East Lansing, MI 48824, USA}
\author{D.~Seymour}
\affiliation{ National Superconducting Cyclotron Laboratory and Department of Physics and Astronomy, Michigan State University, East Lansing, MI 48824, USA}
\author{R.~Shomin}
\affiliation{ National Superconducting Cyclotron Laboratory and Department of Physics and Astronomy, Michigan State University, East Lansing, MI 48824, USA}
\author{W.P.~Tan}
\affiliation{ National Superconducting Cyclotron Laboratory and Department of Physics and Astronomy, Michigan State University, East Lansing, MI 48824, USA}
\author{M.B.~Tsang}
\affiliation{ National Superconducting Cyclotron Laboratory and Department of Physics and Astronomy, Michigan State University, East Lansing, MI 48824, USA}
\author{A.~Wagner}
\affiliation{ National Superconducting Cyclotron Laboratory and Department of Physics and Astronomy, Michigan State University, East Lansing, MI 48824, USA}
\author{H.S.~Xu}
\affiliation{ National Superconducting Cyclotron Laboratory and Department of Physics and Astronomy, Michigan State University, East Lansing, MI 48824, USA}
\author{D.A.~Brown}
\affiliation{Lawrence Livermore National Laboratory, Livermore, CA 94550 USA} 
\author{Y.~Larochelle}
\affiliation{ Department of Chemistry and IUCF, Indiana University, Bloomington, IN 47405, USA}
\author{R.T.~de Souza}
\affiliation{ Department of Chemistry and IUCF, Indiana University, Bloomington, IN 47405, USA}
\author{R.~Yanez}
\affiliation{ Department of Chemistry and IUCF, Indiana University, Bloomington, IN 47405, USA}
\author{R.J.~Charity}
\affiliation{ Department of Chemistry, Washington University, St. Louis, MO 63130, USA}
\author{L.G.~Sobotka} 
\affiliation{ Department of Chemistry, Washington University, St. Louis, MO 63130, USA} 



\begin{abstract}
The interplay of the effects of geometry and collective motion on d-$\alpha$ correlation functions is investigated for central Xe+Au collisions at E/A=50 MeV. The data cannot be explained without collective motion, which could be partly along the beam axis. A semi-quantitative description of the data can be obtained using a Monte-Carlo model, where thermal emission is superimposed on collective motion. Both the emission volume and the competition between the thermal and collective motion influence significantly the shape of the correlation function, motivating new strategies for extending intensity interferometry studies to massive particles.
\end{abstract}

\maketitle

Nuclear collisions provide the only means to study highly excited nuclear matter and its phase transitions under laboratory-controlled conditions. Such collisions produce highly excited nuclear systems that persist momentarily. Space-time information about the fate of these systems can be accessed through the dependence of two-particle correlation functions at low relative momentum on Boson or Fermion symmetries and on the particle mutual nuclear and Coulomb interactions [1-4]. Particularly successful investigations using proton-proton correlations at intermediate energies and pion-pion correlations at high energies have been performed in the last decades [1-8]. Extending these studies to all particle species produced during a reaction represents an important objective because different particles may originate at different stages of the reaction where the densities are different. Correlation functions between complex particles and heavy fragments may be more relevant than light particle correlation functions to stages of the reaction where multifragmentation and the liquid-gas phase transition are expected to occur [8-14]. 

Since thermal velocities decrease with particle mass, $v_{th}\propto\left(m\right)^{-1/2}$, while collective velocities do not depend on mass, the latter become increasingly important as more massive particles are considered [15,16]. This influences the connection between the space-time structures of the system and the measured correlations [15-17] for such particles. In this article, we explore the effects of collective motion on correlation functions constructed with pairs of complex particles interacting by means of the mutual nuclear and Coulomb final state interactions. We present our ideas by studying deuteron-alpha correlation functions measured in $^{129}$Xe+$^{197}$Au collisions at E/A=50 MeV. We demonstrate that the correlation data can be understood only if one incorporates significant effects induced by collective motion and temperature along with the space-time geometry. 

We studied central $^{129}$Xe+$^{197}$Au collisions at E/A=50 MeV, using Au targets of 3 mg/cm$^{2}$ areal density and $^{129}$Xe beams, having an intensity of about 108 pps, from the K1200 cyclotron of the National Superconducting Cyclotron Laboratory at Michigan State University. Isotopically resolved particles with 1$\leq$Z$\leq$10 were detected with nine telescopes of the Large Area Silicon Strip Array (LASSA) [18,19]. Each telescope consisted of one 65 $\mu$m single-sided silicon strip detector, one 500 $\mu$m double-sided silicon strip detector and four 6-cm thick CsI(Tl) scintillators. The strips of the silicon detectors provided an angular resolution of about $\pm$0.43$^{o}$. The center of the LASSA device was located at a polar angle of $\theta$=35$^{o}$ with respect to the beam axis, covering polar angles of 12$^{o}\leq\theta\leq$ 62$^{o}$ and azimuthal angles 24$^{o}\leq\phi\leq$ 156$^{o}$. Impact parameters were selected by the multiplicity of charged particles, measured with LASSA and the 188 plastic scintillator- CsI(Tl) phoswich detectors of the Miniball/Miniwall array [20]; the combined apparatus  covered 80$\%$ 
of the total solid angle. Quasi-central collisions, with reduced impact parameters of the order of b/b$_{max}\leq$0.3, were selected by a gate on the measured charged-particle multiplicity distribution. This reduced impact parameter gate can contain non-negligible contaminations from mid-peripheral events. Statistics requirements characterizing two-particle correlation analyses prevented us from using more selective impact parameter filters as those described in ref. [21].

The data points on Fig. 1 show the d-$\alpha$ correlation function, $1+R(q)$, defined in terms of the two-particle differential multiplicity, $dM_{d\alpha}\left(\vec{p}_{d},\vec{p}_{\alpha}\right)/dp_{d}^{3}dp_{d}^{3}$, and the single particle differential multiplicities, $dM_{d}\left(\vec{p}_{d}\right)/dp_{d}^{3}$ and $dM_{\alpha}\left(\vec{p}_{\alpha}\right)/dp_{\alpha}^{3}$:
\begin{equation}
\Sigma dM_{d\alpha}\left( \vec{p}_{d},\vec{p}_{\alpha}\right)/dp_{d}^{3}dp_{\alpha}^{3}=C\cdot \left[ 1+R\left(q\right)\right]\cdot \Sigma dM_{d}\left( \vec{p}_{d}\right)/dp_{d}^{3}\cdot \Sigma dM_{\alpha}\left( \vec{p}_{\alpha}\right)/dp_{\alpha}^{3}
\label{eq_correl}
\end{equation}
where $\vec{p}_{d}$ and $\vec{p}_{\alpha}$ are the laboratory momenta of the two coincident particles, $q=\mu\cdot v_{rel}$ is the momentum of relative motion, and $C$ is a normalization constant chosen such that $<R(q)>$=0 for large $q$-values where final-state interaction effects are negligible.  The sums on each side of Eq.~(\ref{eq_correl}) extend over all particle energies and angles contributing to each bin of $q$. The product of the single particle yields, on the r.h.s. of Eq.~(\ref{eq_correl}) has been constructed by mixing particles from different events [22]. We apply the same impact parameter selection to both single and two particle multiplicity distributions in Eq.~(\ref{eq_correl}).

The d-$\alpha$ correlation function exhibits a large minimum at small relative momentum due to the mutual Coulomb repulsion. The structures at larger relative momentum arise from the d-$\alpha$ nuclear interaction. In particular, the sharp peak at $q\approx$42~ MeV/c corresponds to the first excited state of $^{6}$Li at E$^{*}$($^{6}$Li) = 2.186 MeV (J$^{\pi}$=3$^{+}$, $\Gamma$=0.024 MeV, $\Gamma_{c}/\Gamma$=1.0) and the broad peak around 84 MeV/c stems mainly from the resonance at E$^{*}$($^{6}$Li) = 4.31 MeV  (J$^{\pi}$=2$^{+}$, $\Gamma$=1.3 MeV, $\Gamma_{c}$/$\Gamma$=0.97) with small contributions from the resonance at E$^{*}$($^{6}$Li) = 5.65 MeV  (J$^{\pi}$=1$^{+}$, $\Gamma$=1.9 MeV, $\Gamma_{c}$/$\Gamma$=0.74). Theoretically, the correlation function is commonly calculated using the angle-averaged Koonin-Pratt equation [1,5-8]:
\begin{equation}
R\left(q,P\right)=4\pi\int dr\cdot r^{2}\cdot K\left(q,r\right)\cdot S\left(r,q,P\right).
\label{eq_kooninpratt}
\end{equation}
The angle-averaged kernel $K(q,r)$ is calculated from the radial part of the d-$\alpha$ elastic scattering relative wave function. The source function, $S(r,q,P)$, is defined as the probability for two-correlated particles with relative momentum $q$ and total momentum $P$ to be separated by a distance $r$ when the last of the two particles is emitted. Most studies using Eq.~(\ref{eq_kooninpratt}) have assumed that the source function depends only on the relative distance, $r$, and not on relative momentum, $q$ [1,2,5-8,23,24]. Under this assumption, Gaussian source profiles, $S(r)\propto \exp\left(-r^{2}/r_{0}^{2}\right)$, depending only on the geometry of the system, have failed in reproducing the line-shape of the correlation functions [2,23,24]. In particular, theoretical correlation functions that reproduce the magnitude of the E$^{*}$=2.186 MeV peak, typically over-predict the E$^{*}\approx$4.31 MeV peak [24]. Even if one accounts for long-lived secondary decays of unstable fragments, the situation does not improve because the relative magnitude of the two peaks does not change. Indeed, as it was shown in the case of proton-proton correlation functions, long-lived secondary decays reduce the magnitude of the correlation peaks without changing their shape [7].  

\begin{figure}
\centering
\includegraphics[scale=0.8]{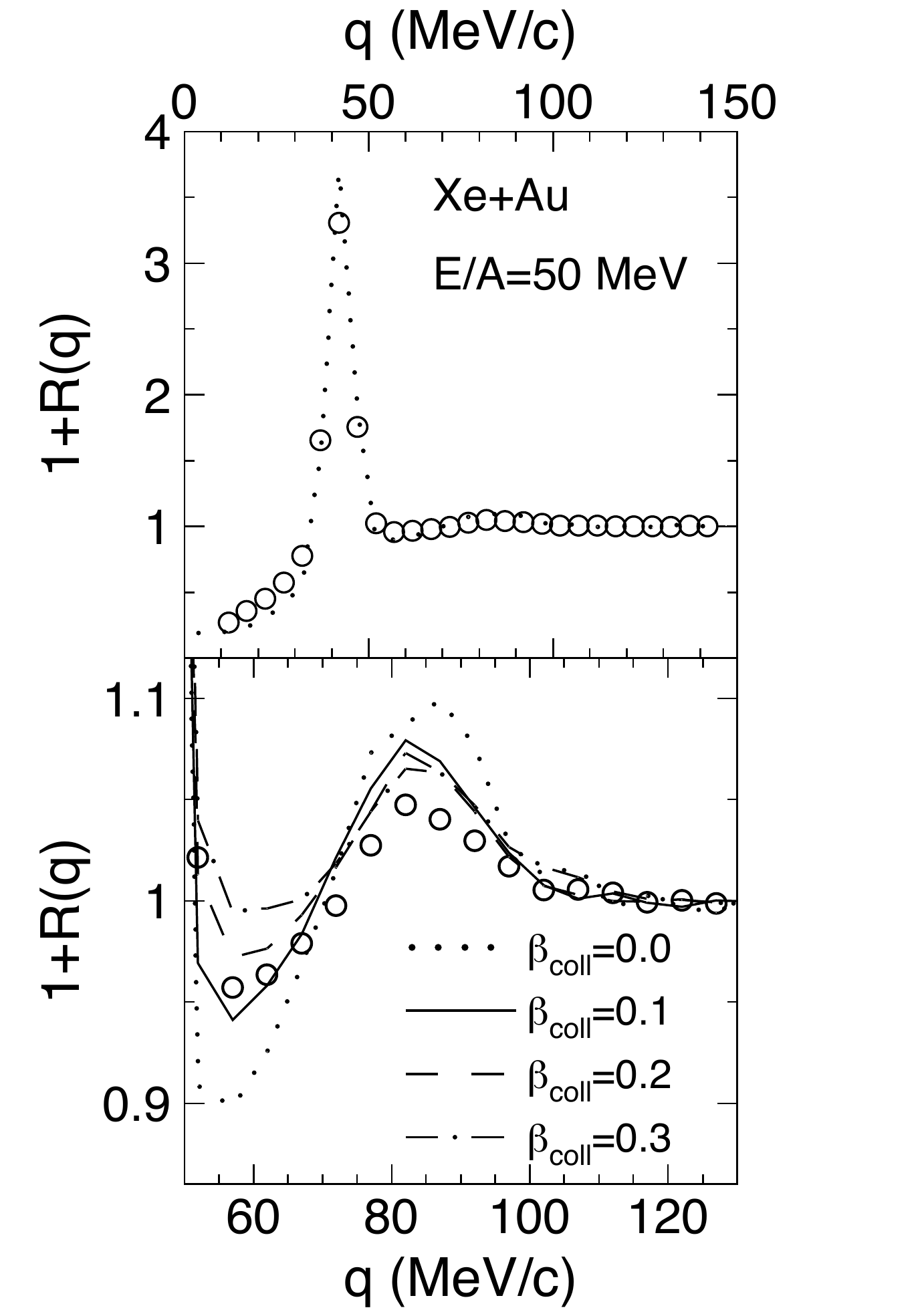}
\caption{d-$\alpha$ correlation function. Top panel: overall view of the correlation function. Bottom panel: view emphasizing the second resonance at $q\approx$84 MeV/c. Circles: d-$\alpha$ correlation function measured in Xe+Au collisions at E/A=50 MeV. Lines: calculated correlation functions for a pure-thermal spherical source with $R_{0}$=10.5 fm, $T$=4 MeV (dotted line), and an ellipsoidal source with ($R_{Z}$=16fm,$R_{X}$=8fm,$R_{Y}$=8fm), $T$=4 MeV and collective motion parameter $\beta_{coll}$=0.1 (solid line), 0.2 (dashed line) and 0.3 (dot-dashed line).}
\label{fig_dacorr}
\end{figure}

Stimulated by the difficulties encountered by previous approaches, we explore in details how the line-shape of a d-$\alpha$ correlation function can be affected by the interplay of the geometry of the source with thermal and collective motion. We introduce a Monte Carlo simulation where deuterons and alpha particles are emitted at random positions $\vec{r}_{d}$ and $\vec{r}_{\alpha}$ within a spherical single-particle source having radius $R_{0}$. This source evaporates particles while undergoing collective expansion. Then the velocities of the two emitted particles consist of a thermal and a collective component, $\vec{v}_{d}=\vec{v}_{d,th}+\vec{v}_{d,coll}\left(\vec{r}_{d}\right)$ and $\vec{v}_{\alpha}=\vec{v}_{\alpha,th}+\vec{v}_{\alpha,coll}\left(\vec{r}_{\alpha}\right)$. The thermal velocities, $\vec{v}_{d,th}$ and $\vec{v}_{\alpha,th}$, are randomly assigned according to Boltzmann distributions corresponding to an assigned temperature, $T$=4 MeV, consistent with measurements of isotopic and excited states thermometers [25]. The collective velocity components are parameterized as vector fields, $\vec{v}_{d,coll}\left(\vec{r}_{d}\right)$ and $\vec{v}_{\alpha,coll}\left(\vec{r}_{\alpha}\right)$, depending on particle positions, $\vec{r}_{d}$ and $\vec{r}_{\alpha}$. Simulated particle trajectories are filtered through the acceptance of the LASSA array. Finally, the relative distance, $\vec{r}=\vec{r}_{d}-\vec{r}_{\alpha}$, and the momentum of relative motion, $\vec{q}=\mu\left(\vec{v}_{d}-\vec{v}_{\alpha}\right)$, are evaluated and used to construct the two-particle emitting source, $S(r,q)$, by integrating over total momentum, $P=\left|\vec{p}_{d}+\vec{p}_{\alpha}\right|$. The source function is then used in Eq.~(\ref{eq_kooninpratt}) to calculate the correlation function, $R(q)$.

We first compare our data to calculations performed with a pure-thermal emitting source where particle velocities have only their thermal components (corresponding to the assigned temperature, $T$=4 MeV) and no collective components ($\vec{v}_{d,coll}=\vec{v}_{\alpha,coll}$=0). The main impact of varying the radius of the source, $R_{0}$, is to change the widths of the peaks [7]. The width of the first sharp peak at $q$=42 MeV/c is dominated by the finite resolution of the LASSA array, leaving no sensitivity to the source size. In contrast, the width of the broad peak at 84 MeV/c is found to scale almost linearly with the radius of the source, i.e. $\Delta q_{FWHM}$(MeV/c)$\approx$25-1.2$\cdot R_{0}$(fm). This sensitivity provides the opportunity to determine the source size from measured d-$\alpha$ correlation functions. Calculations with a spherical source with $R_{0}$=10.5 fm approximately reproduce the width of the peak at $q\approx$84 MeV/c. As for the p-p correlation function, the main effect of long-lived secondary decays is to reduce the magnitude of the correlation peaks without changing their shape [7]. We take this into account by fitting the measured correlation function $R_{exp}(q)$ by $R_{fast}(q)=\lambda\cdot R(q)$, where $\lambda$ represents the fraction of d-$\lambda$ pairs from the fast source and $R(q)$ is calculated from Eq.~(\ref{eq_kooninpratt}). The dotted line on Fig. 1 shows the calculated correlation function obtained by choosing $\lambda$=0.8. Choosing this value of the $\lambda$ parameter normalizes the correlation function to the first peak at 42 MeV/c, but over-predicts the height of the second peak at $q\approx$84 MeV/c and under-predicts the valley between the two peaks, as seen in the lower panel of Fig. 1 showing an expanded view of $R(q)$ for 50$\leq q\leq$130. These difficulties, already encountered in previous works and confirmed by our high resolution data, cannot be overcome by changing $R_{0}$, $\lambda$, the source temperature or by altering the source geometry. Indeed, the d-$\alpha$ kernel function, $K(r,q)$, has the same spatial extension close to the resonances at $q\approx$42 and 84 MeV/c: any change in the geometry of the source would change the height of the two peaks by the same factor and leave their ratio unmodified. In the following, we focus our attention upon this long-standing and well documented [24] problem and defer optimizations with respect to source size, temperature and geometry to later investigations. 

In order to explain the observed discrepancies, we recognize that our selection of events with high measured charged-particle multiplicities with $\hat b\leq0.3$  includes events with non-negligible collective motion, both along and perpendicular to the beam axis [21,26,27]. For simplicity we parameterize collective velocities as $\vec{v}_{d,coll}\left(\vec{r}_{d}\right)=c\beta_{coll}\vec{r}_{d}/R_{0}$ and $\vec{v}_{\alpha,coll}\left(\vec{r}_{\alpha}\right)=c\beta_{coll}\vec{r}_{\alpha}/R_{0}$ , where $c$ is the velocity of light and $\beta_{coll}$ represents the maximum collective velocity of a particle emitted at the surface of the system, i.e. when $r_{d}=r_{\alpha}=R_{0}$. This parameterization recalls self-similar radial expansion profiles already used in the literature [21,26-29]. 

Choosing deuteron-alpha particle pairs of from the simulation that are detected in LASSA, we construct the source function $S(r)$ that describes their relative separation at emission and found that collective motion induces an apparent reduction of the source size, consistent with studies at higher incident energies [16,17]. The thick solid line in Fig. 2 shows the profile of $S(r)$ for a source with $\beta_{coll}$=0.2, $R_{0}$=10.5 fm and $T$=4 MeV; the dotted line shows $S(r)$ for a pure thermal source with $\beta_{coll}$=0, $R_{0}$=10.5 fm and $T$=4 MeV. In this pure-thermal case, particles have velocities that are independent of the position where they are emitted. Since particles emitted from any location in the source have an equal probability of being detected by the LASSA array, the corresponding correlation function probes the whole spatial extent of the source. If collective motion is present, however, particles emitted from regions where the collective motion points away from LASSA will be included in the correlation function only if their thermal velocity is larger than the local collective velocity. This limits the portion of the emitting source that can be probed by the experimental data and the source size appears reduced. In the particular case shown on Fig. 2, the apparent source size is reduced by as much as 50$\%$.

\begin{figure}
\centering
\includegraphics[scale=0.8]{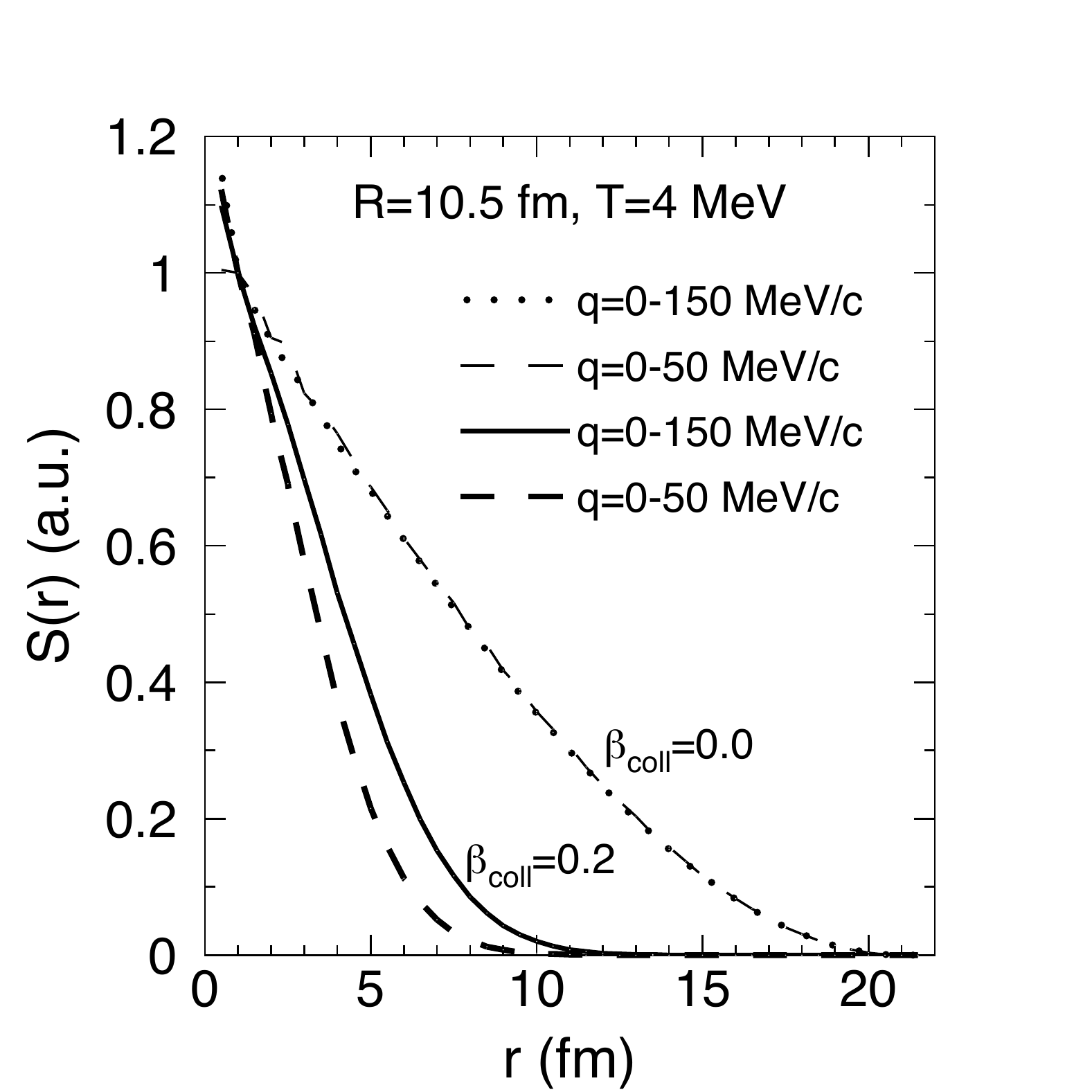}
\caption{Calculated d-$\alpha$ sources for a spherical single-particle source with $R_{0}$=10.5 fm and $T$=4 MeV. Pure-thermal source calculations, $\beta_{coll}$=0.0, are represented by a dotted line (all $q$Õs) and by a thin dashed line ($q$=0-50 MeV/c). The thick solid and thick dashed lines represent the spherical source with $\beta_{coll}$=0.2 and, respectively, without and with a $q$=0-50 MeV/c gate applied.}
\label{fig_dasou}
\end{figure}

The source size reduction is just one aspect of the influence of collective motion on correlations. Another is a dependence of the correlation function on the total momentum,  $\vec{P}=\vec{p}_{1}+\vec{p}_{2}$ of the particle pairs [1,2-8,16,17]. Such total momentum-dependence has also been interpreted at intermediate energies as a consequence of emission from cooling and expanding systems [1-4,23]. In Fig. 2 we show that the extracted source size for particles within a small relative momentum gate, $q$=0-50 MeV/c, is even smaller (thick dashed line) than the size of source corresponding to all extracted d-$\alpha$ pairs (solid line). Thus, collective motion makes the source function both $r$- and $q$-dependent. It happens because the d-$\alpha$ relative momentum, $\vec{q}$, has thermal and collective components, $\vec{q}=\vec{q}_{th}+\vec{q}_{coll}\left(\vec{r}\right)$. The collective component, given by $\vec{q}_{coll}\left(\vec{r}\right)=\mu\frac{c\beta_{coll}}{R_{0}}\vec{r}$, depends on the relative distance, making the source both $r$- and $q$-dependent, i.e. $S=S(r,q)$, with small relative momentum values being  positively correlated with correspondingly small relative distances. This correlation between relative distance, $r$, and relative momentum, $q$, is an unexplored effect of collective motion on two-particle emitting sources. In the case of a pure-thermal model ($\beta_{coll}$=0), the relative momentum is independent of $r$ and contains only a thermal component, $\vec{q}\approx\vec{q}_{th}=\mu\left(\vec{v}_{d,th}-\vec{v}_{\alpha,th}\right)$. Indeed, in Fig. 2 we show that the pure-thermal source profile obtained by gating at small relative momentum, $q$=0-50 MeV/c (thin dashed line) is exactly the same as the original ungated one  (dotted line). 

To show how collective motion influences the shape of the d-$\alpha$ correlation function, we separate it into two components, $R(q)=R_{C}(q)+R_{N}(q)$, where $R_{C}(q)$ and $R_{N}(q)$ correspond to the correlation functions generated separately by the mutual d-$\alpha$ Coulomb and nuclear interactions, respectively. The left and the right panels of Fig. 3 show, respectively, $1+R_{C}(q)$ and $1+R_{N}(q)$ obtained for a source with $R_{0}$=10.5 fm, $T$=4 MeV, and $\beta_{coll}$=0.0 (dotted line), $\beta_{coll}$=0.1 (solid line), $\beta_{coll}$=0.2 (dashed line) and $\beta_{coll}$=0.3 (dot-dashed line). The nuclear correlation functions (right panel)  are all normalized to the integral of the first peak at $q$=42 MeV/c, not shown on Fig. 3.

\begin{figure}
\centering
\includegraphics[scale=0.8]{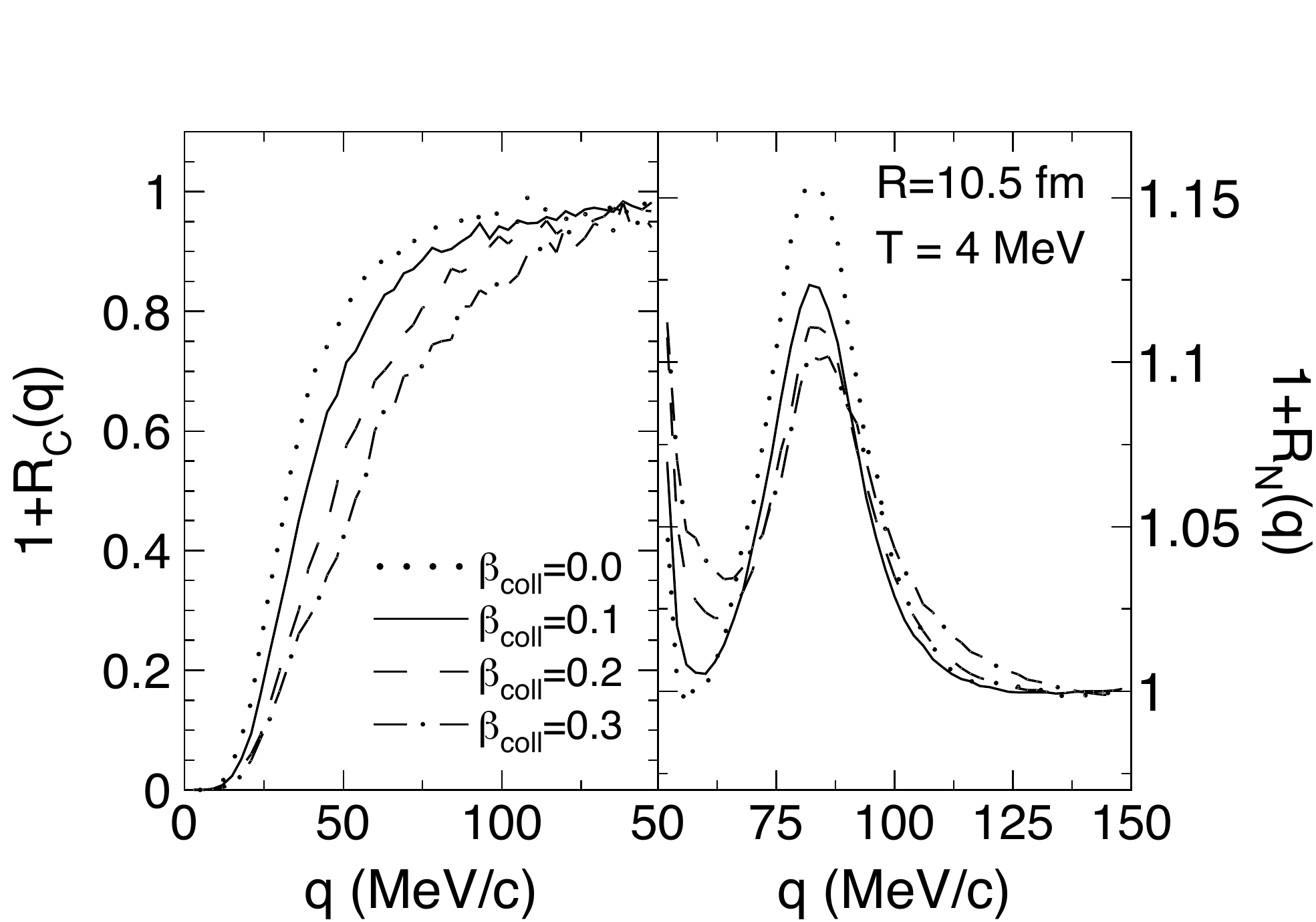}
\caption{Effects of collective motion on Coulomb (left panel) and Nuclear (right panel) correlation functions. Dotted, solid, dashed and dot-dashed lines correspond to calculations performed, respectively, with $\beta_{coll}$=0.0, $\beta_{coll}$=0.1, $\beta_{coll}$=0.2 and $\beta_{coll}$=0.3.} 
\label{fig_danuclcoul}
\end{figure}

The Coulomb correlation function (left panel, Fig. 3) displays one prominent anti-correlation at small relative momentum due to the mutual Coulomb repulsion between the deuteron and the alpha particle. The width of the Coulomb dip increases with collective velocity, consistent with a decreasing apparent source size caused by the collective motion in the system. This result shows that the possible existence of collective motion must be taken into account in order to extract source sizes and emission times from Coulomb dominated correlation functions [30].

On the right panel of Fig. 3 we show the calculated nuclear correlation functions in the region where the second peak is observed ($q$=50-110 MeV/c). The magnitude of the peak at 84 MeV/c decreases with increasing $\beta_{coll}$. In order to understand this effect, we point out that the relative momentum regions where resonant peaks exist, are mostly dominated by d-$\alpha$ pairs emitted at very small relative distances, $r\leq$2 fm, corresponding to the average d-$\alpha$ spatial separation in the resonance. Under these conditions, the two coincident particles have nearly the same collective velocities because they originate from nearly the same location, resulting in a collective component of relative momentum given by $\vec{q}_{coll}\left(r\approx 0\right)\approx 0$ and $q\approx q_{th}$. Then, the relative momentum distribution can be described as a Boltzman distribution at the local temperature, $T$, of the source. The uncorrelated two-particle spectrum in the denominator of Eq.~(\ref{eq_correl}), on the other hand, is calculated by selecting deuterons and alpha particles detected in different events. Either particle can originate from anywhere in the source, thus the collective velocities of d and $\alpha$ can be very different and $\vec{q}_{coll}\ne 0$. Then the uncorrelated relative momentum spectrum is nearly exponential but has a larger slope, $T_{MIX}>T$, as compared to that of the coincidence two-particle spectrum. The resulting correlation function, obtained by folding the coincidence spectrum with the kernel $K(r,q)$ and by dividing by the uncorrelated spectrum, displays an exponential fall-off in those relative momentum regions where resonances are observed. The slope, $T_{app}$, of such fall-off is approximately given by the relation $1/T_{app}=1/T-1/T_{MIX}$ and it is the origin of the attenuation of the height of the second peak observed in the right panel of Fig. 3. This effect becomes increasingly important as the masses of the particles are increased, reflecting the more relevant role played by collective motion in the case of heavier particles. Previous studies with complex particles have commonly assumed that the temperature dependence in the numerator and denominator of Eq.~(\ref{eq_correl}) are the same and cancel out in calculating their ratio [23]. As a consequence, the resulting correlation function resulted sensitive only to the volume of the source and not to the temperature [23]. In contrast, we show on Fig. 3 that even small amounts of collective motion make the effective temperature, $T_{MIX}$, of the denominator larger that the thermal temperature. The resulting correlation function is then sensitive not only to the volume of the source but also to its temperature through the interplay of geometry with collective motion. Our result provides the basis for unified approaches to correlation function analyses to extract both space-time and thermal properties of emitting sources. 

To compare calculated correlation functions to our data in Fig. 1, we make more realistic assumptions about the geometry and the collective motion of the emitting source. Transverse collective expansions of about 2 MeV per nucleon have been previously observed in central Xe+Sn symmetric collisions at E/A=50 MeV by studying complex fragment kinetic energy spectra [21]. It has also been observed that for central events selected by high measured charged particle multiplicities, the longitudinal collective motion in more asymmetric Xe+Au reactions is larger [31]. Thus, we simulate the emission of deuterons and alpha particles to occur from an ellipsoidal source characterized by radii $R_{Z}=2R_{X}=2R_{Y}$, where $Z$ indicates the beam direction. Deuterons and $\alpha$ particle collective velocities are described as $\vec{v}_{d,coll}\left(\vec{r}_{d}\right)=c\beta_{coll}\left(\vec{r}_{d}/R_{Z}\right)$ and $\vec{v}_{\alpha,coll}\left(\vec{r}_{\alpha}\right)=c\beta_{coll}\left(\vec{r}_{\alpha}/R_{Z}\right)$, respectively. This parameterization simulates the expected radial collective motion in the center of mass of the reaction and provides collective velocity components to those particles emitted close to the edges ($r_{d}$,$r_{\alpha}\approx R_{Z}$) of the ellipsoid, representing the effects of incomplete stopping of the incident momenta.

Correlations functions have been calculated for a ranges of radii, $(R_{Z},R_{X},R_{Y})$, and for different values of the radial velocity parameter, $\beta_{coll}$=0.1, 0.2 and 0.3. We searched for those combinations of $(R_{Z},R_{X},R_{Y})$ and $\beta_{coll}$ that provided a width of the peak at 84 MeV/c in reasonable agreement with the data. A reasonable description of that width is obtained for $R_{Z}$=15-20 fm, $R_{X}$=7.5-10 fm, $R_{Y}$=7.5-10 fm and $\beta_{coll}\approx$0.1-0.2. In the bottom panel of Fig. 1, we only show the results corresponding to $R_{Z}$=16 fm, $R_{X}$=$R_{Y}$=8 fm. The different curves correspond to $\beta_{coll}$=0.1 (solid line), $\beta_{coll}$=0.2 (dashed line) and $\beta_{coll}$=0.3 (dot-dashed line). All simulated correlation functions shown in Fig. 1 are normalized to reproduce the magnitude of the first peak at $q$=42 MeV/c. Values of $\beta_{coll}$=0.1-0.2, providing a btter agreement with the data, induce expansion velocities in the central region of the order of $\beta_{rad}\approx \beta_{coll}/2\approx$ 0.05-0.1, consistent with previous measurements of transverse flow for Xe+Sn and Xe+Au collisions [21,31]. Maximum collective velocities at the edges of the source ($r=R_{Z}$) are of the order of $\beta_{coll}\approx$0.1-0.2, consistent with the velocity of projectile-like and target-like remnants in the center of mass of the studied reactions. By introducing collective motion (solid line), we find that one can reduce significantly the observed discrepancies between experimental data and the pure-thermal source calculation (dotted line). The height of the peak at 84 MeV/c still remains slightly over-predicted. The valley between the peak at 42 and 84 MeV/c is well described by the calculations. More refined descriptions of collective motion (expansion, rotation, etc.) and of the break-up geometry might provide somewhat better agreement. However, our main goal is to demonstrate that the interplay between thermal and collective motion introduces a $q$-dependence in the source function that can account for the long-standing difficulties in describing the line-shape of d-$\alpha$ correlation functions. Given the range of radii that yield reasonable descriptions of the data and assuming that the source contains about 80$\%$ 
of the total mass [21,31], we find that the source volume is consistent with average nuclear densities ranging between $\rho\approx$0.2$\rho_{0}$ and 0.4$\rho_{0}$ (with $\rho_{0}$=0.16 fm$^{-3}$, saturation density). Such range is consistent with the freeze-out density values commonly employed in statistical models of multifragmentation [32].

The present work is limited to the study of inclusive d-$\alpha$ pairs, without any condition imposed on their velocities or total momentum. Such inclusive data contain contributions from different stages of the reaction. The existence of these multiple sources of deuteron and $\alpha$ particles has been predicted by the Expanding Emitting Source model [33] and is expected to influence the shape of correlation functions [7]. We expect that a quantitative comparison to experimental data can be improved by implementing our simplified single-source model within a more realistic scenario where multiple sources with different timescales are included. Such ideas can be also tested by gating on the total momentum of the correlated particles. 

In summary, in this work we show that the interplay between source geometry and collective motion plays a key role in determining the line-shape of correlation functions constructed with complex particles. In addition to the previously studied effects induced by collective motion, we observe also the presence of a significant relative momentum dependence in the emitting source. We apply our ideas to study a d-$\alpha$ correlation function measured in Xe+Au collisions at E/A=50 MeV. Realistic assumptions about the volume, the temperature and the collective motion in the system allow us to obtain a semi-quantitative description of the data and explain the difficulties encountered in previous studies. The presented results also show, for the first time, that the detailed line-shape of high resolution correlation functions probe not only the space-time geometry of the emitting source but also its interplay with thermal and collective motion. 
 
\begin{acknowledgments}
This work is supported by the National Science Foundation under Grant Nos. PHY-0245009 and PHY-0555893. Part of this work was performed under the auspices of the U.S. Department of Energy by University of California, Lawrence Livermore National Laboratory under Contract W-7405-Eng-48. 
\end{acknowledgments}



\begin{thebibliography}{99}

\bibitem{bib1}S.E. Koonin, Phys. Lett. B70, 43 (1977)
\bibitem{bib2} D.H. Boal, C.K. Gelbke and B.K. Jennings, Rev. Mod. Phys. 62, 553 (1990)
\bibitem{bib3} U. Heinz and B.V. Jacak, Ann. Rev. Nucl. Part. Sci. 49, 529 (1999)
10
\bibitem{bib4} M. Lisa, S. Pratt, R. Soltz and U. Wiedemann, Ann. Rev. Nucl. Part. Sci. 55, 357 (2005)
\bibitem{bib5} D.A. Brown and P. Danielewicz, Phys. Lett. B398, 252 (1997).
\bibitem{bib6} D.A. Brown and P. Danielewicz, Phys. Rev. C64, 104902 (2001)
\bibitem{bib7} G. Verde et al., Phys. Rev. C 65, 054609 (2002).
\bibitem{bib8} D.A. Brown and P. Danielewicz, Phys. Rev. C 57, 2474 (1998).
\bibitem{bib9} M. D'Agostino et al., Phys. Lett. B371, 175 (1996).
\bibitem{bib10} O. Schapiro, A.R. DeAngelis, and D.H.E. Gross, Nucl. Phys. A568, 333 (1994).
\bibitem{bib11} R. Popescu et al., Phys. Rev. C 58, 270 (1998)
\bibitem{bib12} L. Beaulieu et al., Phys. Rev. Lett. 84, 5791 (2000)
\bibitem{bib13} J. Pochodzalla et al., Phys. Rev. Lett. 75, 1040 (1995)
\bibitem{bib14} J. Natowitz et al., Phys. Rev. C 65, 034618 (2002)
\bibitem{bib15} R.T. de Souza et al., Physics Letters B 300, 29 (1993)
\bibitem{bib16} R. Kotte et al., Phys. Rev. C 51, 2686Ð2699 (1995).
\bibitem{bib17} K. Adkox et al., Phys. Rev. Lett. 88, 192302 (2002).
\bibitem{bib18} B. Davin et al., Nucl. Instr. and Meth. A 473, 302 (2001).
\bibitem{bib19} A. Wagner et al., Nucl. Instr. and Meth. A456, 290 (2001)
\bibitem{bib20} R.T. de Souza et al., Nucl. Inst. Meth. A 295, 109 (1990).
\bibitem{bib21} N. Marie et al., Phys. Lett. B391, 15 (1997)
\bibitem{bib22} M.A. Lisa et al., Phys. Rev. C44, 2865 (1991).
\bibitem{bib23} J. Pochodzalla et al., Phys. Rev. C35, 1695 (1987).
\bibitem{bib24} C.B. Chitwood et al., Phys. Lett. B172, 27 (1986).
\bibitem{bib25} H.F. Xi et al., Phys. Rev. C58, R2636 (1998)
\bibitem{bib26} D.R. Bowman et al., Phys. Rev. C52, 818 (1995)
\bibitem{bib27} W. Reisdorf and H.G. Ritter, Annu. Rev. Nucl. Part. Sci. 47, 663 (1997)
\bibitem{bib28} W.C. Hsi et al., Phys. Rev. Lett. 73, 3367 ~1994
\bibitem{bib29} M.A. Lisa et al., Phys. Rev. Lett. 75, 2662 ~1995
\bibitem{bib30} E. Cornell et al, Phys. Rev. Lett. 75, 1475 (1995)
\bibitem{bib31} C. Williams et al., Phys. Rev. C55, R2132
\bibitem{bib32} W.P. Tan et al., Phys. Rev. C 68, 034609 (2003)
\bibitem{bib33} G.J. Kunde et al., Phys. Lett. B416, 56 (1998)

\end{thebibliography}

\end{document}